\documentclass[twocolumn]{aastex63}

\usepackage{amsmath}
\usepackage{graphicx}
\usepackage{dcolumn}
\usepackage{bm}
\usepackage{xcolor}

\newcommand{\dHybridR}{{\it dHybridR}}
\newcommand{\ocii}{\Omega_{ci}^{-1}}
\newcommand{\di}{d_{i}}
\newcommand{\va}{v_{A,0}}
\newcommand{\w}[1]{v_{A,#1}}
\newcommand{\vcr}{v_{\rm{CR},x}}
\newcommand{\vup}{M\va}
\newcommand{\vsh}{v_{\rm sh}}
\newcommand{\rt}{R_{\rm tot}}
\newcommand{\rs}{R_{\rm sub}}

\newcommand{\pinj}{p_{\rm inj}}
\newcommand{\teff}{\vartheta_{\rm eff}}

\submitjournal{ApJ}

\shorttitle{CR-Modified Shocks I: Hydrodynamics}
\shortauthors{Haggerty \& Caprioli}

\graphicspath{{./}{figures/}}

\begin{document}

\title{Kinetic Simulations of Cosmic-Ray–Modified Shocks I: Hydrodynamics}

\correspondingauthor{Colby C. Haggerty}
\email{chaggerty@uchicago.edu}

\author[0000-0002-2160-7288]{Colby C. Haggerty}
\author[0000-0003-0939-8775]{Damiano Caprioli}
\affiliation{Department of Astronomy and Astrophysics, University of Chicago, 5640 S Ellis Ave, Chicago, IL 60637, USA}

\date{\today}

\begin{abstract}
Collisionless plasma shocks are efficient sources of non-thermal particle acceleration in space and astrophysical systems. 
We use hybrid (kinetic ions -- fluid electrons) simulations to examine the non-linear feedback of the self-generated energetic particles (cosmic rays, CRs) on the shock hydrodynamics.
When CR acceleration is efficient, we find evidence of both an upstream precursor, where the inflowing plasma is compressed and heated, and a downstream postcursor, where the energy flux in CRs and amplified magnetic fields play a dynamical role.
For the first time, we assess how non-linear magnetic fluctuations in the postcursor preferentially travel away from the shock at roughly the local Alfv\'en speed with respect to the downstream plasma.
The drift of both magnetic and CR energy with respect to the thermal plasma substantially increases the shock compression ratio with respect to the standard prediction, in particular exceeding 4 for strong shocks.
Such modifications also have implications for the spectrum of the particles accelerated via diffusive shock acceleration, a significant result detailed in a companion paper, \cite{caprioli+20}, \url{https://arxiv.org/abs/2009.00007}.

\end{abstract}

\section{\label{sec:intro}Introduction}
Non-relativistic shocks are abundant in space and astrophysical systems, such as the Earth's bow shock, interplanetary shocks associated with coronal mass ejections, supernovae (SNe) and supernova remnants (SNRs), and galaxy clusters;
they are typically associated with non-thermal particles and emission.
Such shocks occur on length scales where collisions are too infrequent to efficiently dissipate the energy flux;
in fact, they are referred to as collisionless shocks because the energy dissipation is regulated by the collective interactions between the charged particles and the electromagnetic fields.
Despite this, much of our understanding and predictions for the macroscopic structure of plasma shocks comes from the (collisional) fluid Rankine--Hugoniot jump conditions, which cannot accurately model the dynamics of non-thermal particles and magnetic fields.
Non-relativistic shocks can transfer as much as 10\% -- 20\% of their ram energy into energetic particles (hereafter, cosmic rays, CRs), through repeated first-order Fermi reflections, a process referred to as diffusive shock acceleration \citep[DSA,][]{krymskii77,axford+78,bell78a,blandford+78}.
Additionally, CRs drive plasma instabilities upstream of the shock, which  amplify magnetic fluctuations and produce turbulent, large-amplitude magnetic fields, which are further compressed in the downstream region \citep[e.g.,][]{skilling75a,bell04,bykov+13,caprioli+13,caprioli+14b}.

Previous theoretical works have included CRs in modeling the hydrodynamics of collisionless plasma shocks using two-fluid theory \citep[e.g.,][]{drury-volk81a, drury-volk81b, berezhko+99}, analytical kinetic theory \citep[e.g.,][]{eichler79,eichler84,eichler85,ellison+85b,malkov97,MDV00,blasi02,blasi04,amato+05,amato+06,caprioli+09a,caprioli+10a,caprioli+09b}, and Monte Carlo numerical approaches \citep[e.g.,][]{ellison+84,ellison+85b,ellison+90,ellison+95,ellison+96,ellison+02,vladimirov+06}. 
Further numerical approaches have also been employed to understand the time dependence of this effect on different astrophysical shocks \citep[e.g.,][]{bell87,jones+91,berezhko+97,berezhko+99,giesler+00,berezhko+04a,volk+05,berezhko+06a,kang+02,kang+05,kang+06,zirakashvili+08b,zirakashvili+10,zirakashvili+08,kang+13}. 
For detailed reviews of CR-modified shocks, readers can refer to \cite{drury83,blandford+87, jones+91, malkov+01}.

When even a relatively small fraction of shock energy ($\sim 10\%$) is channeled into CRs, an upstream precursor is formed and the speed and compressibility of the shock are affected at the zero-th order.
The standard prediction is that the CR contribution should lead to a larger total shock compression ratio, possibly as large as 10-100 for strong shocks \citep[e.g.][]{drury83,jones+01,malkov+01}.
The effect of self-generated, amplified magnetic fields has been predicted to limit such a compression to values $\lesssim 10$ \citep{vladimirov+06,caprioli+08, caprioli+09b}, but always greater than the fiducial value of 4 for shocks with Mach number $\gg 1$.
While there is general agreement that self-generated CRs and magnetic fluctuations should modify the hydrodynamics of a shock, the process through which this occurs has not been assessed from first principles, yet.

In this paper we present the first numerical evidence of CR-modified shocks in {\it ab-initio} plasma simulations in the hybrid limit (kinetic ions -- fluid electrons).
We detail how CRs and CR-driven magnetic fluctuations affect the hydrodynamics of the shock, while in a companion paper, \cite{caprioli+20}, we discuss how the modified hydrodynamics affect the spectrum of the particles accelerated via DSA.

The layout of this paper is as follows: in \S\ref{sec:sims}, we detail the hybrid code and the shock simulation setup.
In \S\ref{sec:pre} and \S\ref{sec:post}, we identify and discuss the formation of a precursor region in the upstream and a postcursor region in the downstream, respectively. 
In \S\ref{sec:hydro}, we present CR modified shock jump conditions and show they are in good agreement with simulations.
Finally, in \S\ref{sec:disc}, we discuss the implication of these corrections to the fluid dynamics of collisionless plasma shocks.

\section{\label{sec:sims}Hybrid Simulations}
To study the non-linear effects of self-generated CRs on the hydrodynamics of the shock, we perform self-consistent simulations using \dHybridR{}, a relativistic hybrid code with kinetic ions and (massless, charge-neutralizing) fluid electrons \citep{haggerty+19a}.
\dHybridR{} is the generalization of the Newtonian code \emph{dHybrid} \citep{gargate+07}, which has been already widely used for simulating collisionless shocks \citep{gargate+12,caprioli+14a,caprioli+14b,caprioli+14c,caprioli+15, caprioli+17, caprioli+18, haggerty+19p,caprioli+19p}.
Hybrid codes are better suited to self-consistently simulate the long-term shock evolution than full particle-in-cell codes because they do not need to resolve the small time/length scales of the electrons, which are usually dynamically negligible  \citep[see, e.g.,][and references therein]{winske85,quest88,scholer90,giacalone+92,giacalone+93,bennett+95,winske+96,giacalone+97,giacalone+00,giacalone04,burgess+05,lipatov02,guo+13,burgess+16,kropotina+16,hanusch+19}. 

All physical quantities are normalized to their far upstream values, namely: mass density to $\rho_0 \equiv m_i n_0$ (with $m_i$ the ion, namely proton, mass), magnetic fields to $B_0$, lengths to the ion inertial length $\di = c/\omega_{pi}$ (with $c$ the speed of light and $\omega_{pi}$ the ion plasma frequency), time to the inverse ion cyclotron frequency $\ocii$, and velocity to the Alfv\'en speed $\w0 = B_0/\sqrt{4\pi \rho_0}$.
The ion temperature is chosen such that the thermal gyroradius is $1\, \di$, corresponding to an ion thermal to magnetic pressure ratio of $\beta_i = 2$.
The system is 2D in real space (in the $x-y$ plane), and all the three components of momenta and electromagnetic fields are retained.
The hybrid model requires an explicit choice for the electron equation of state, and in this work, electrons are assumed to be adiabatic, i.e., the electron pressure is $P_e\propto\rho^{5/3}$.
We remark on the motivation and consequences of choosing this equation of state in \S\ref{sec:disc}.
\begin{figure}[t]
\includegraphics[width=.48\textwidth,clip=true,trim= 0 0 0 0]{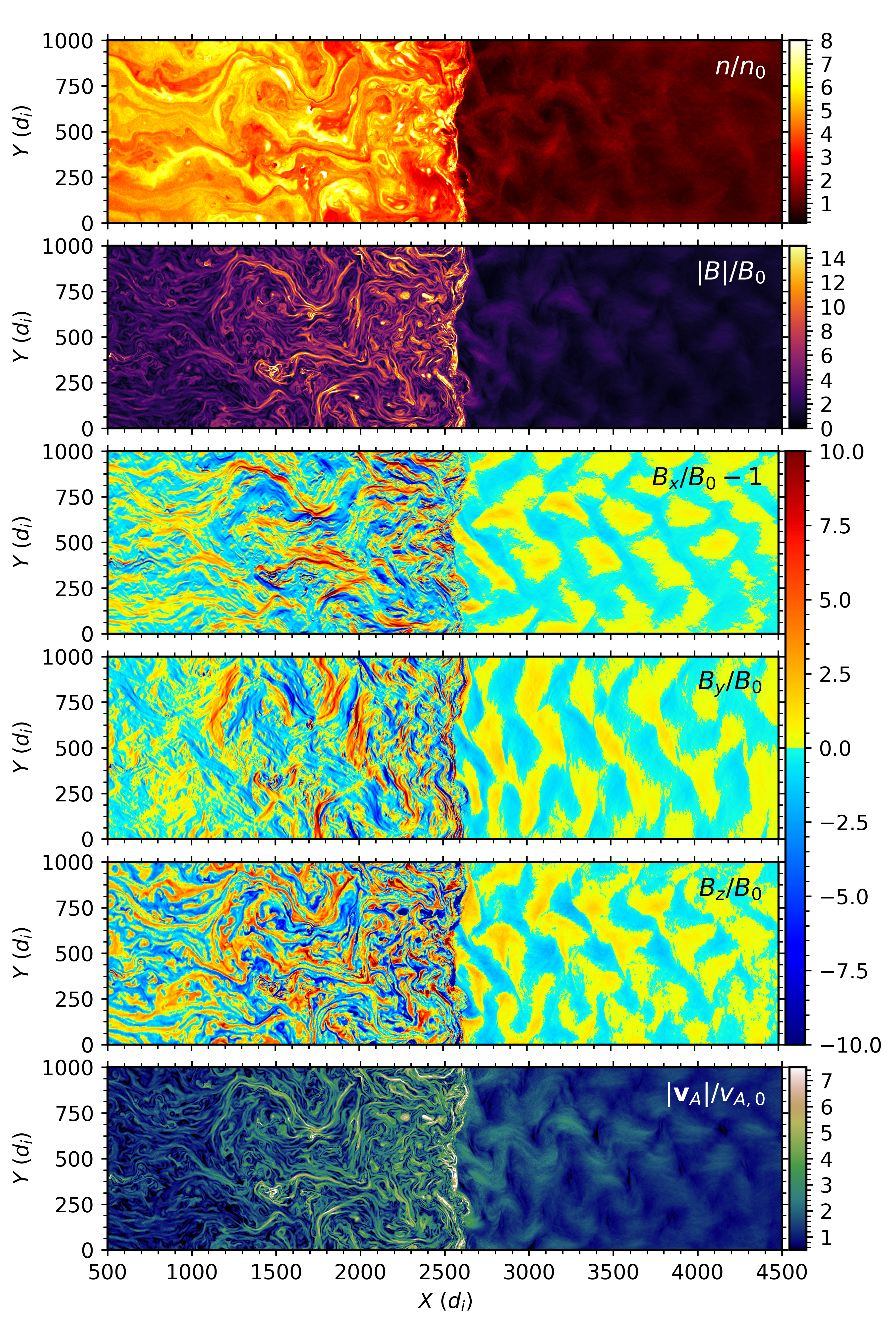}
\caption{2D snapshot of a $M=20$ parallel shock at time $t = 500 \ocii$. Quantities are (top to bottom):  density, magnitude of the magnetic field, self-generated components of the magnetic field ($B_x/B_0-1$, $B_y$ and $B_z$) and magnitude of the local Alfv\'en speed $|{\bf v}_A| = |{\bf B}|/\sqrt{4\pi m_i n}$. 
The shock is at $x\sim 2500 \di$.}\label{fig:overview}
\end{figure}

The simulations are initialized with a uniform magnetic field ${\bf B_0}=B_0 {\bf x}$ and with a thermal ion population with a bulk flow ${\bf v}_x = - M \va{\bf x}$). 
The simulations are periodic in the $y$ direction, the right boundary is open and continuously injecting thermal particles, and the left boundary is a reflecting wall;
after tens of cyclotron times, the ion population closest to the wall reflects, becomes unstable, and forms a shock.
This shock then travels in the $+{\bf x}$ direction, parallel to the background magnetic field (\emph{parallel shock}), with the downstream plasma at rest in the simulation reference frame.
Note that $M$ defines both sonic and Alfv\'enic Mach numbers $M \equiv M_A = \sqrt{\gamma}M_s$ in the downstream (simulation) frame, while the Mach number that enters the stationary jump conditions is the one in the shock frame, which is 20-30\% larger, depending on the shock compression ratio (see \S\ref{sec:hydro}).
Most of the analysis in this work is done on a benchmark simulation with $M = 20$ and a domain of size $[L_x,\, L_y] = [10^5, 200] \di$, wide enough to account for 2D effects and long enough so that the simulation could be run for more than $1000\, \ocii$ without energetic particles escaping the box.
The simulation has two grid cells per $\di$, and each grid cell is initialized with four particles per grid.
The speed of light is set to be much larger than the Alfv\'en and thermal speeds ($c/\w0 = c/v_{thi} = 100$), as discussed in \cite{haggerty+19a}; the time step is set as $ c\Delta t = d_i/10$. 

Simulations with Mach numbers in the range $M=10 - 80$ were performed;
their parameters are detailed in Appendix A. 
All of the following analysis and figures in this work are preformed with the $M=20$ benchmark simulation, unless stated otherwise.  
The choice stems from the fact that a $M=20$ simulation is representative of a strong shock ($M \gg 1$) and can still be run for a long time without CRs escaping the simulation domain; 
the computational cost of a \dHybridR{} simulation scales $\propto M^2$ when keeping the box size fixed in units of the ion gyroradius (for a Newtonian hybrid code it would scale $\propto M^3$, since also $dt$ is inversely proportional to $M$ if not normalized to $c$.) 

Moreover, for $M = 20$ the fastest growing modes, driven by streaming CRs, are in the resonant regime \citep[non-resonant modes start to become prominent for $M\gtrsim 30$, see][]{caprioli+14b};
this means that the amplified fields should be Alfv\'enic and remain quasi-linear upstream, which simplifies the theoretical interpretation.
We have also performed runs with larger $M$ (\S\ref{sec:Mach}), for which magnetic field amplification occurs in the Bell regime, finding consistent results for all strong shocks.  

A shorter $M=20$ simulation with a wider box $L_y = 10^3 \di$  was run for assessing convergence with the transverse size; quantities from this simulation are shown in Figure \ref{fig:overview}.
With these parameters, strong parallel shocks can channel as much as $10 - 15\%$ of their kinetic energy into CRs and effectively amplify magnetic fields \citep{caprioli+14a, caprioli+14b, caprioli+14c}.

\section{\label{sec:pre}The Upstream Precursor}
\subsection{Hydrodynamics}
\begin{figure}[t]
\includegraphics[width=.48\textwidth,clip=true,trim= 0 0 0 0]{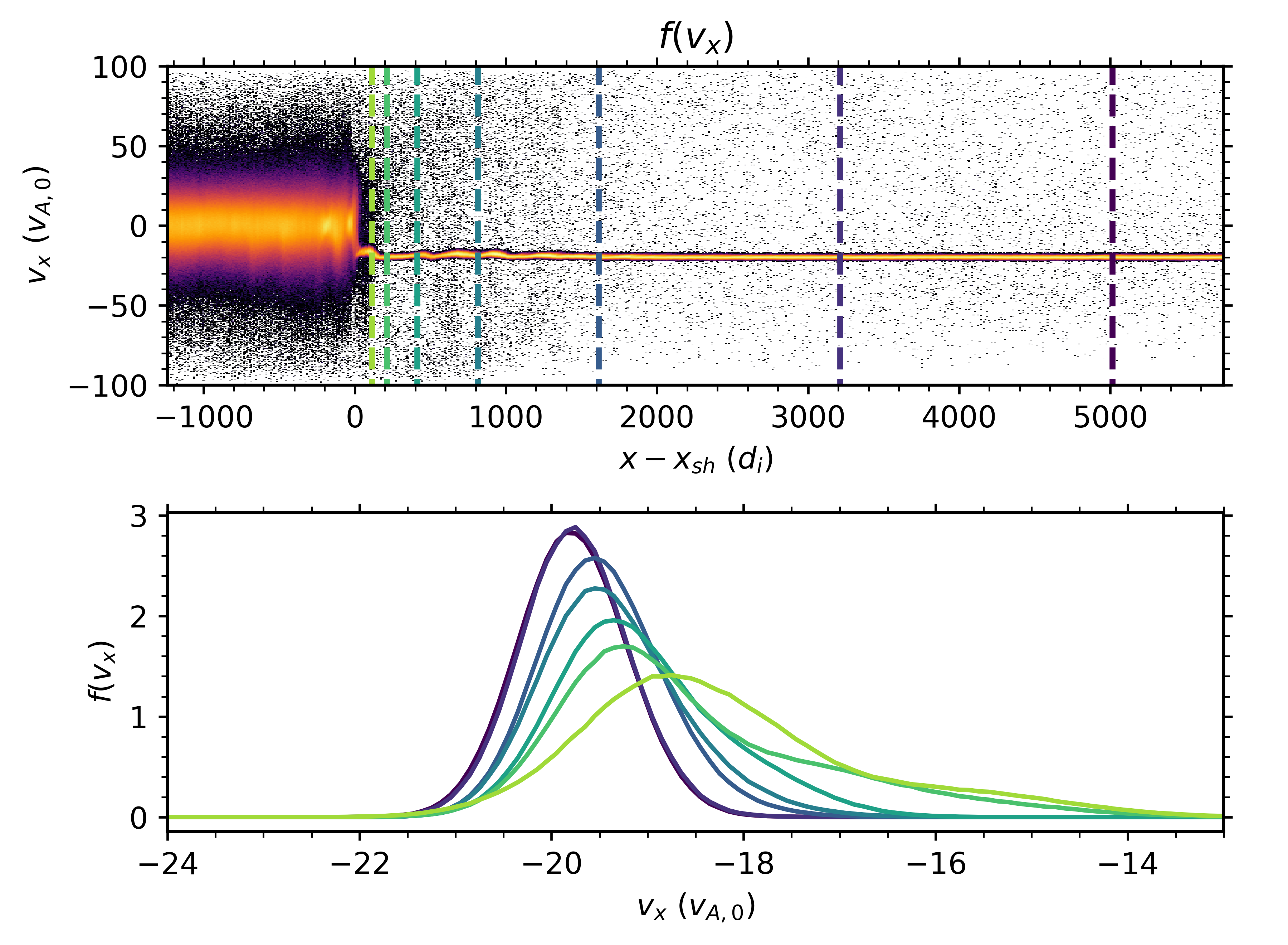}
\caption{Top panel: ion $x-v_x$ phase space distribution function at $t =1400 \ocii$, with the $x$ axis shifted to the shock is at $x=0$
Bottom panel: Ion spectra at different upstream positions (color coded as in the top panel); 
cuts are time-averaged between $t = 1350 - 1550\ocii$ to remove high frequency fluctuations. 
The bulk speed decreases and the thermal speed increases while approaching the shock.}
\label{fig:precursor}
\end{figure}
When a large enough fraction of the shock kinetic energy and pressure are deposited into energetic particles, the standard shock hydrodynamics are altered from the ones described by the gaseous Rankine-Hugoniot jump conditions.
CRs break the causality wall represented by the shock, and thus, can transfer momentum and energy back upstream, effectively slowing the shock and pre-compressing the incoming plasma, forming a CR-induced \emph{precursor} \citep[e.g.,][]{drury83,jones+91,blandford+87,malkov+01}. 
For the quantities measured in this work, we use the subscripts 0, 1 and 2  for the far upstream, the precursor immediately upstream of the shock, and downstream respectively.

The formation of the CR precursor is evident in our simulations:
the upper panel of Figure \ref{fig:precursor} shows the $x-v_x$ phase space distribution function integrated/averaged over the $y, v_y$ and $v_z$ directions at $1400 \ocii$. Cuts of the ion $v_x$ distribution are shown in the lower panel. Each cut is taken at different distances from the shock (color coded) and averaged over $t = 1350 - 1550 \ocii$.
The bulk flow speed (i.e., the mean of the $v_x$ distribution) upstream of the subshock is $u_1\sim 0.9 u_0$, consistent with pressure in CRs being $\sim 10\%$ of the bulk ram pressure.
In the precursor, thermal ions warm up (i.e., the $v_x$ distributions become wider) well beyond what is predicted by adiabatic heating due to the observed compression $\rho_1/\rho_0\sim 1.09$\footnote{Mass flux conservation implies that $\rho u$ is constant.}.
A temperature increase consistent with adiabatic heating would yield $T_{1}/T_{0} = (\rho_{1}/\rho_{0})^{2/3} \approx 1.06$, while Figure \ref{fig:precursor} indicates a much larger temperature increase, $T_{1}/T_{0}\sim 2.7$.
Such a non-adiabatic heating is due to the damping of the waves produced by CR-driven instabilities \citep[e.g.,][]{ellison+81,volk+81, amato+06,caprioli+09a,tatischeff+07} and corresponds to maintaining a constant thermal/magnetic pressure ratio in the shock precursor, despite the effective magnetic field amplification. 

\subsection{Magnetic Field Amplification}
\begin{figure}[t]
\includegraphics[width=.45\textwidth,clip=true,trim= 0 0 0 0]{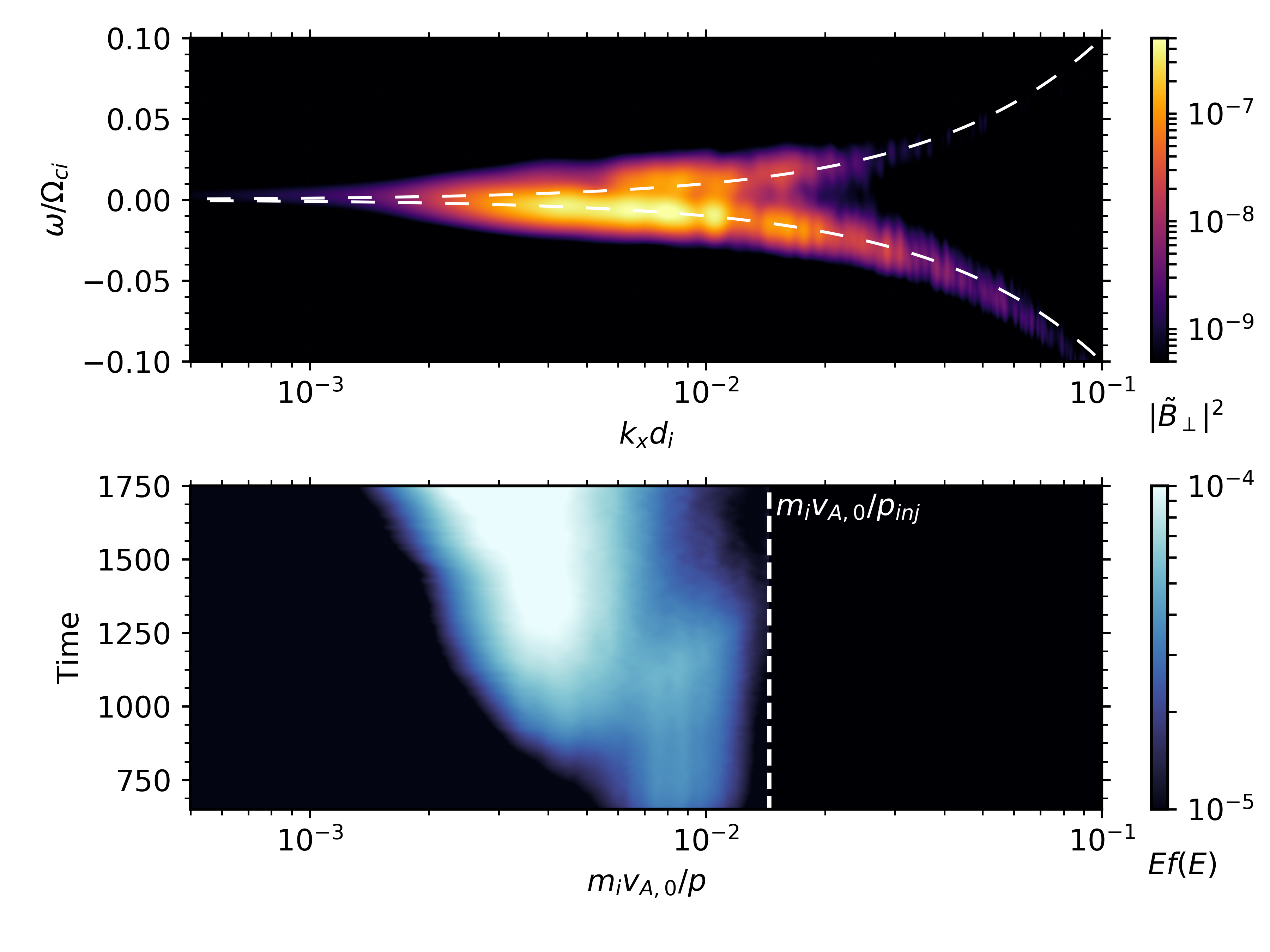}
\caption{Top Panel: Power in the upstream transverse magnetic field components $|\Tilde{B}_\perp|^2 \equiv |\Tilde{B}_y|^2+|\Tilde{B}_z|^2$, where $\Tilde{B_i}$ is the Fourier transform of $B_i(x,t)$, as a function of wave number ($k$) and angular frequency ($\omega$).
In this plot, modes with positive(negative) $\omega$ correspond to waves traveling left(right) and dashed white lines to Alfv\'en phase velocity of $\pm\w0$.
Self-generated modes are Alfv\'enic and preferentially move to the right, i.e., against the fluid.
Bottom Panel: The upstream CR energy spectrum, plotted in time (vertical axis) as a function of $m_i\w0/p$, which can be roughly interpreted as the resonant wave number.
The comparison between panels highlights the quasi-resonance between the CR and wave power spectra.}\label{fig:kw_up}
\end{figure}

The CR population is intrinsically anisotropic in the upstream fluid frame, which leads to the development of several plasma instabilities that  have been investigated theoretically and numerically in numerous works \citep[e.g.,][]{kulsrud+69,skilling75a,bell78a,zweibel03,bell04,reville+08a,amato+09,riquelme+09,reville+13,caprioli+14b,bai+19,haggerty+19p,zacharegkas+19p}.
The presented kinetic simulations capture self-consistently both the formation of the precursor and the CR-driven magnetic fluctuations.

The power spectrum of self-generated magnetic fluctuations in the upstream is shown in Figure \ref{fig:kw_up} for our benchmark run. 
The spectrum is calculated using the transverse components of the magnetic field with $|\Tilde{B}_\perp|^2 \equiv |\Tilde{B}_y|^2+|\Tilde{B}_z|^2$, where $\Tilde{B_i}$ is the Fourier transform of $ B_i(x,t)$, as a function of wave number $k$ and angular frequency $\omega$\footnote{In practice, we calculate the discreet Fourier transform of $B_i(x,y,t)$ in the $x$ and $t$ directions, average over the y direction, and normalize to the number of grid points in $x$ and $t$}.
Also, it is evaluated over a relatively large window [$60000\, \di$,\ $200\, \di$], embedded in the upstream flow, with the left side of the window beginning at $x = 23600\, \di$ at $t = 600\, \ocii$ and ending at $x = 8000\di$, just in front of the shock, at $t = 1780\,  \ocii$.
The phase velocity of an Alfv\'en wave, based on the initial upstream magnetic field and density is shown by the white dashed lines, with the upper and lower lines corresponding to waves traveling  towards (leftward) and away from (rightward) the shock respectively.
Most of the magnetic power is in modes with phase speeds comparable to the initial upstream Alfv\'en speed and with $\omega < 0$, i.e., modes that are propagating \emph{away from the shock} in the fluid rest frame and down the CR pressure gradient.

Furthermore, most of the power is in modes with wave numbers between $kd_i \sim 0.002 - 0.02$, which are resonant with the CR population upstream of the shock.
The resonant CR population can be seen in the bottom panel of Figure \ref{fig:kw_up}, which shows a 2D map of the ion energy spectrum, $Ef(E)$, integrated over the transformation window. The spectrum is plotted as a function of time and inverse momentum ($m_i\w0/p$), which, for a constant magnetic field strength, can be interpreted as the corresponding resonant wave number. 
Note that the CR spectrum is cut off for values smaller than the injection momentum ($\pinj,$ vertical dashed line), so that the signal is not overpowered by the inflowing beam of ions with $p = 20m_i\w0$ \citep{caprioli+15}.
By comparing the x-axis of the energy spectrum and magnetic power spectrum, it can be concluded that, for the benchmark simulation, the amplified modes are driven by the resonant CR streaming instability, \citep[e.g.,][]{kulsrud+69,skilling75a,bell78a,zweibel03};
this result is in agreement with theoretical expectations \citep{amato+09} and previous hybrid simulations \citep{caprioli+14b}, which suggest that the non-resonant (Bell) instability should become dominant only for $M\gtrsim 30$.

\subsection{The Quasi-periodic Nature of the Precursor}
\begin{figure}[t]
\centering
\includegraphics[width=.47\textwidth,clip=true,trim= 0 0 0 0]{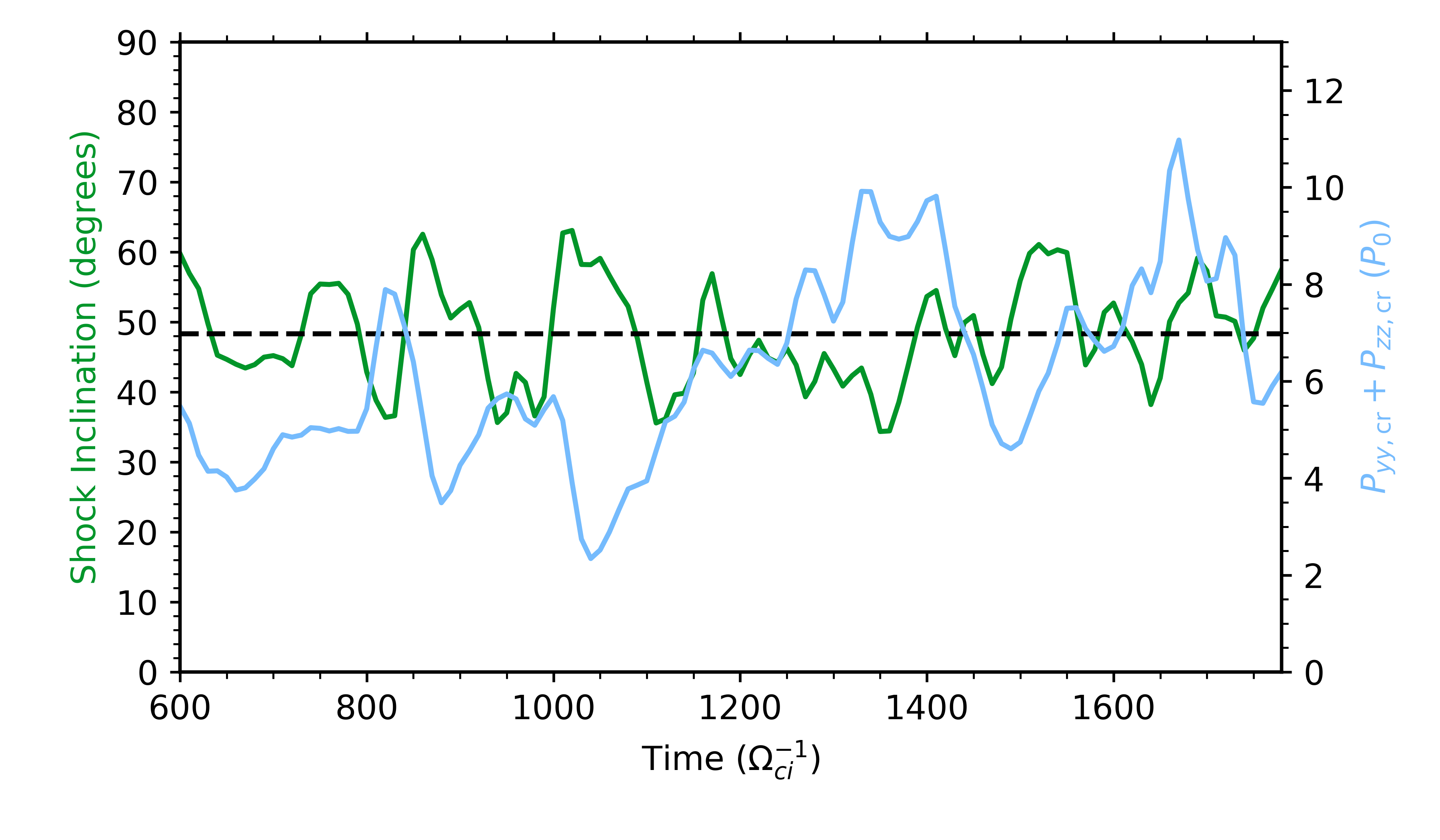}
\caption{Angle between the magnetic field immediately upstream of the shock and the shock normal ($\teff=\cos^{-1}(\left < B_{x,1}/|B_1|\right >,$ left axis, green line), and total transverse CR pressure in the upstream ($P_{yy} + P_{zz}$, right axis, light blue line, normalized to $P_0 \equiv \rho_0v_{A,0}^2$).
Note the quasi-periodic behaviour, with a time-scale of $\sim 200\ocii$ and the general anticorrelation between shock inclination and CR pressure.}\label{fig:shock_inc}
\end{figure}

While the standard predictions for the precursor are recovered in time-averaged simulation data, a high-cadence analysis shows that the precursor varies significantly on intermediate time scales, around $50 - 200 \ocii$.
These time scales roughly match the period of the Doppler-shifted Alfv\'enic fluctuations in the precursor\footnote{Most of the power in resonant modes are at wavelengths of $\lambda \sim 2\pi/k \sim 2000\, \di$ (Figure \ref{fig:kw_up}).
 With a bulk flow of $\sim 20\w0$, the period of these oscillations in the shock frame are $\sim \lambda/(20\w0) \sim 100\ocii$.}.
 
A global time variability originates from the CR-driven instabilities and the spatially inhomogeneous amplification of the upstream magnetic field. 
In fact, the effective shock inclination, $\teff(t)$, is modulated on time-scales comparable to the wave period. Furthermore, the shock inclination controls particle injection into DSA, which is more prominent for quasi-parallel shock configurations and suppressed for oblique ($\teff\gtrsim 50\deg$) ones.

This modulation can be seen in Figure \ref{fig:shock_inc}, where the dark green line (left axis) shows $\teff(t)\equiv\cos^{-1}(\left <B_x/|B| \right>)$, with the magnetic field averaged over a region between 350 and 450 $\di$ upstream of the shock.
Sufficiently far upstream the field is mostly along its initial direction, but $\teff(t)$ oscillates around its average of $\approx 48^\circ$ (dashed line) with a period comparable to that of the self-generated waves.

Figure \ref{fig:shock_inc} also shows the transverse CR pressure, defined as $P_{yy} + P_{zz}$, where $P_{ii} = \int_{\pinj}^{\infty} p_iv_i f d^3p$ in units of the upstream thermal pressure $P_0$ (light blue, right axis), calculated over the same region as $\teff$.
Such a CR pressure varies with a period comparable to the shock inclination too, but the two quantities appear to be generally anti-correlated.
We conclude that CR injection, the prominence of the precursor, and the shock inclination are all modulated by the period of the waves generated via CR streaming instability in the precursor.
The imprint of upstream magnetic fluctuations on the shock dynamics is expected to survive also in more realistic (much wider and 3D) systems, though as a local phenomenon that, arguably, causes shock rippling.
This quasi-periodic nature of precursors may be measurable {\it in-situ} at heliospheric shocks and must be reckoned with when determining the fundamental inclination of a shock based on an instantaneous measurement  $\teff$.

\section{\label{sec:post}The Downstream Postcursor}
\begin{figure}[ht]
\includegraphics[width=.48\textwidth,clip=true,trim= 0 0 0 0]{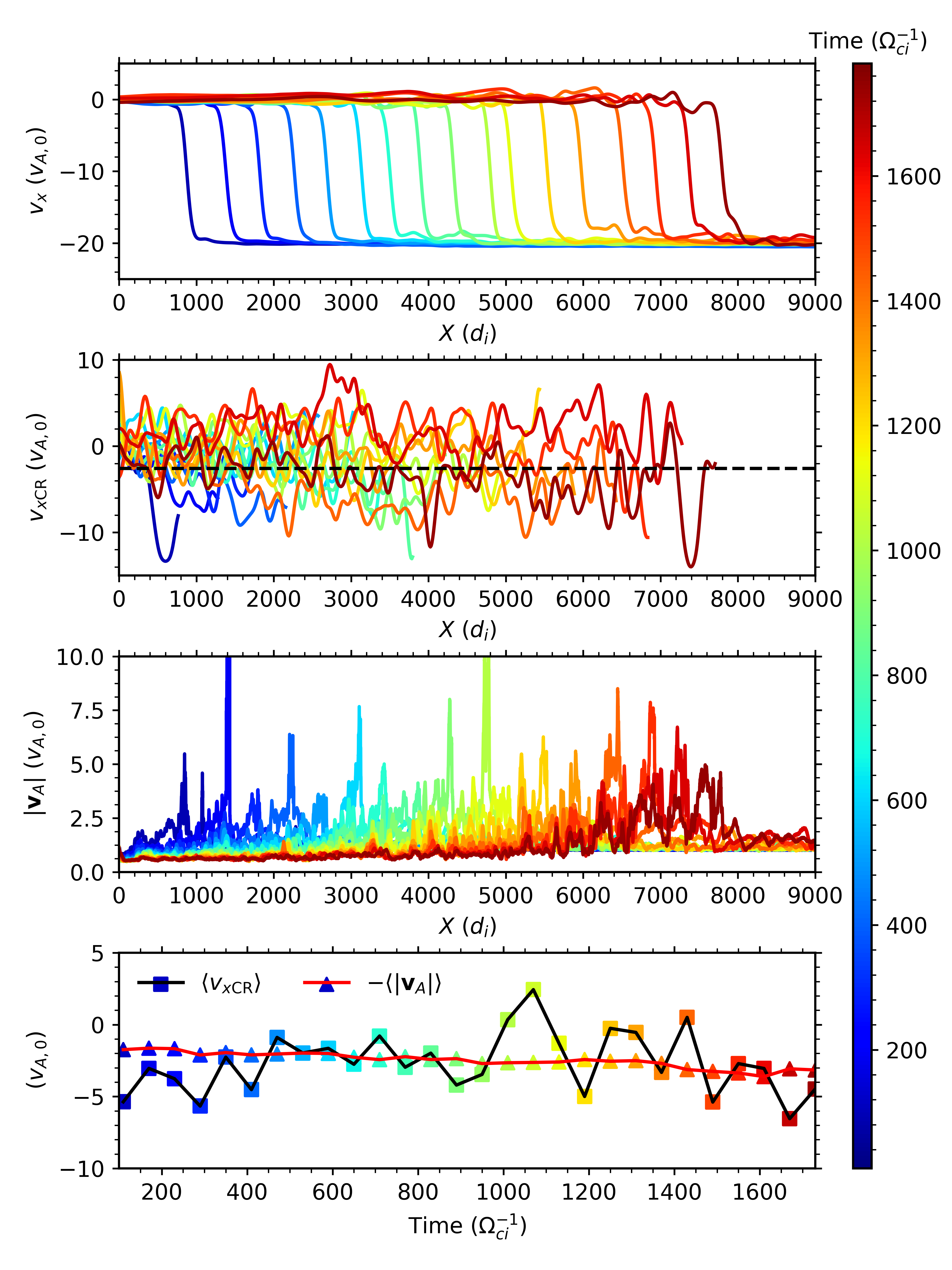}
\caption{Time evolution of the velocity profiles, averaged over the transverse direction, for: all the ions ($v_x$), CR only ($\vcr$, see Equation \ref{eq:vcr}),
and local Alfv\'en speed, $|v_A| = |{\bf B}|/\sqrt{4\pi \rho}$ (top three panels).
The bottom panel compares CR (black line, squares) and Alfv\'en (red line, triangles) speeds averaged $500 \di$ behind the shock, suggesting that CRs drift towards the downstream with respect to the thermal plasma at speeds comparable to the local magnetic fluctuations.
The horizontal dashed lines mark the time-averaged postcursor Alfv\'en speed.
}\label{fig:alpha}
\end{figure}
In contrast to the formation of the shock precursor, what was not predicted by any theory (that we are aware of) and what we found for the first time in self-consistent plasma simulations is the formation of a non-linear structure downstream of the shock, which we call a \emph{postcursor}.

The postcursor manifests itself as \emph{an extended downstream region where the self-generated magnetic fluctuations have an important dynamical role and the dynamics of thermal plasma and CRs are decoupled, in the sense that CRs and magnetic perturbations have a sizable drift speed with respect to the bulk plasma}.

A standard assumption in DSA theory is that CRs are quickly isotropized in the frame of the magnetic fluctuations, usually assumed to be Alfv\'en waves \citep{skilling75a}, and that the Alfv\'en speed is much smaller than the flow speed in the shock frame, so that eventually CRs are isotropic in the flow frame at the order of $v_A/u$.
Deviations from such an isotropy may arise if the shock is oblique and trans-relativistic \citep[e.g.,][]{baring+95,ellison+96,bell+11} or in general in the presence of anisotropic transport \citep[e.g.,][]{kirk+96,morlino+07a}.

In the presence of magnetic field amplification, the condition $v_A/u\ll 1$ may be violated: it has been suggested \citep{zirakashvili+08b, caprioli+09a,caprioli11,caprioli12} that retaining the \emph{upstream} Alfv\'enic drift of CRs with respect to the background fluid may also affect the resulting CR spectrum  appreciably  \citep{morlino+12,kang+13,slane+14,kang+18,bell+19}.
Retaining this effect upstream is natural, since self-generated waves travel against the CR gradient, i.e., towards upstream infinity resulting in a net CR drift $\vcr\simeq u-v_A$, as shown in the top panel of Figure~\ref{fig:kw_up}. 
However, it has always been assumed that \emph{downstream} magnetic fluctuations should not have a preferential direction, hence canceling any net drift \citep[see][for a rigorous derivation of CR transport in the presence of waves of both helicities]{skilling75a}.

\subsection{CR Drift Downstream}
In this work we find, for the first time, evidence for a net CR drift away from the shock, as shown in Figure \ref{fig:alpha}.
In all the panels, the color code corresponds to time evolution.
The top two panels show the bulk velocity normal to the shock, for all the ions (thermal + CRs, $v_x$, top panel) and for the CRs alone ($\vcr$ bottom panel). The speeds are averaged along $y$ and shown as a function of $x$. 
The CR velocity is defined as the mean velocity of particles with momentum larger than $\pinj$, i.e.,
\begin{equation}\label{eq:vcr}
        \left. {\bf v}_{\rm CR} \equiv \int_{\pinj}^\infty {\bf v}p^2 f dp 
        \middle/ \int_{\pinj}^\infty p^2f dp \right.,
\end{equation}
where $\pinj \approx \sqrt{10}mu_0$ according to the injection theory derived from hybrid simulations in \cite{caprioli+14a, caprioli+15}.

While the downstream plasma is --as expected-- at rest behind the shock in the simulation frame (top panel of Figure \ref{fig:alpha}), the velocity of the CR population has a net negative value of about $2-3 \w0$ in the postcursor, about $500\di$ wide just downstream, and reduces in magnitude further downstream until vanishing close to the left wall (second panel).
Ultimately, the drift of CRs is controlled by their interaction with the magnetic fluctuations, and so, this flux is expected to be tied to the dynamics of the downstream magnetic field.

The third panel of Figure \ref{fig:alpha} shows the local Alfv\'en speed $|{\bf v_A}(x)| = \left <|{\bf B}(x,y)|/\sqrt{4\pi\rho(x,y)}\right >_y$, averaged in the $y$ direction (see also the bottom panel of Figure \ref{fig:overview}).
It is important to stress that this quantity is only informative about the typical speed of the magnetic perturbations, which are not simple Alfv\'en waves but rather large-amplitude structures created by CR-driven instabilities in the precursor and further compressed at the shock. 
In the paper we will commonly use ``wave'' for brevity, bearing in mind that magnetic structures are often nonlinear.
Finally, the bottom panel of Figure \ref{fig:alpha} compares directly the average CR drift velocity and the postcursor Alfv\'en speed: they are consistently similar in time and approach $|{\bf v_A}| \approx 3 \w0$ at later times.

In summary, in the postcursor, the  downstream CRs drift away from the shock with respect to the thermal plasma at a speed comparable to the local Alfv\'en speed;
the extent of the CR postcursor is determined by the spatial extent of the amplified magnetic field.

\subsection{Motion of Post-shock Magnetic Structures}
\begin{figure*}
\centering
\includegraphics[width=.9\textwidth,clip=true,trim= 0 5 4 5]{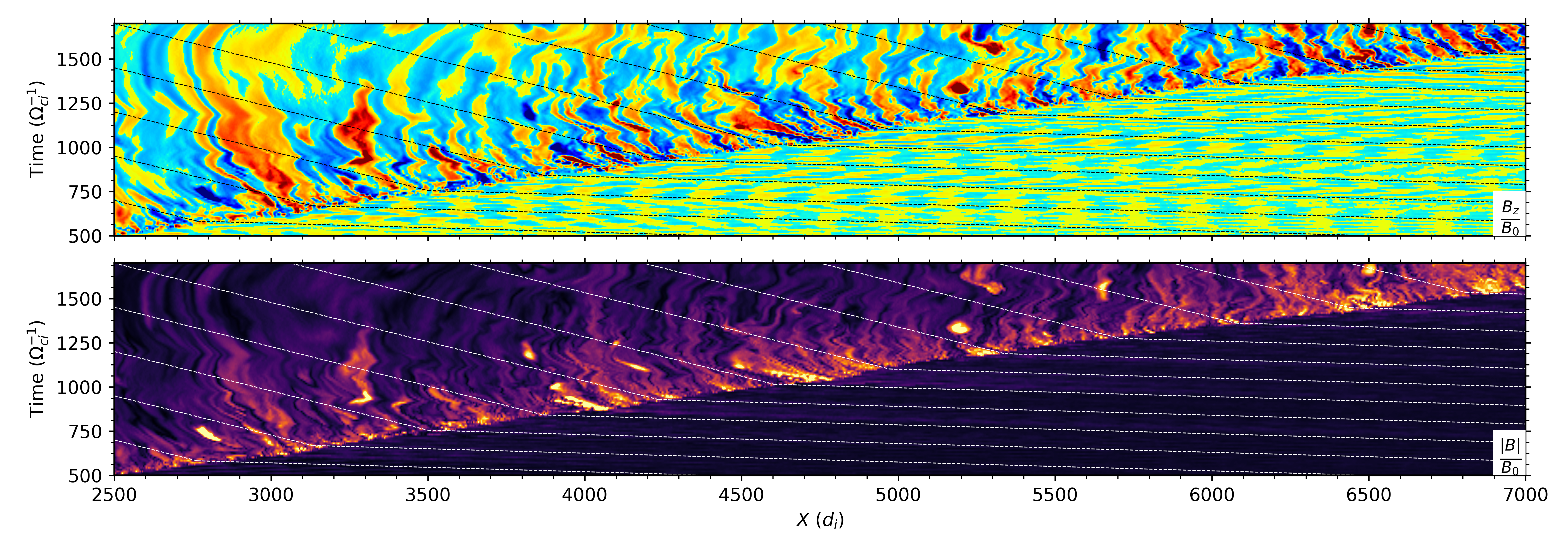}
\caption{Evolution of the simulation magnetic field. 
At any given time, a 1D cut is plotted (averaged on 5$\di$ around $y=50\di$) for $B_z$ and $|{\bf B}|$ (left and right).
The diagonal dashed lines correspond to average bulk flow + Alfv\'en speed in the upstream and postcursor.
Immediately behind the shock, the magnetic structures align with such lines, which means that the phase motion of such structures points away from the shock; further downstream, the structures are more vertical, i.e., at rest in the fluid frame.
}\label{fig:Bxt}
\end{figure*}

Even if there is agreement in the magnitude of CR drift and Alfv\'en speeds, we are left with the task to demonstrate that the frame in which CRs are isotropic is actually the wave frame.
Note that in simulations we observe CRs drift \emph{away from the shock} relative to the thermal plasma, and it is not obvious that  magnetic fluctuations should have a preferred direction of propagation downstream.
To quantify this, we present two distinct diagnostics: a morphological analysis of the magnetic structures behind the shock and a calculation of the wave dispersion relations.

The evolution of the magnetic field ($B_z$ and $|{\bf B}|$) is shown in Figure \ref{fig:Bxt}: 1D cuts are taken along the $x$ direction, averaged between $y = \{47.5, 52.5\}\, \di$, and plotted from $x = 2500\di$ to $7000 \di$ over the last $1200\ocii$ of the simulation.
In each panel the shock front makes a nearly straight line with a positive slope corresponding to the inverse of the shock speed;
the downstream region lies above this line.
The inclination of the post-shock, finger-like, magnetic structures provides an estimate of their velocity in the simulation frame: just behind the shock, waves have a negative velocity consistent with dashed lines of $-2.5\w0$, while further downstream they are almost at rest (i.e., vertical).
This suggests that, in the postcursor, magnetic fluctuations are traveling \emph{away} from the shock faster than the thermal plasma, consistent with the net CR drift.

\begin{figure}[ht]
\includegraphics[width=.45\textwidth,clip=true,trim= 5 0 10 0]{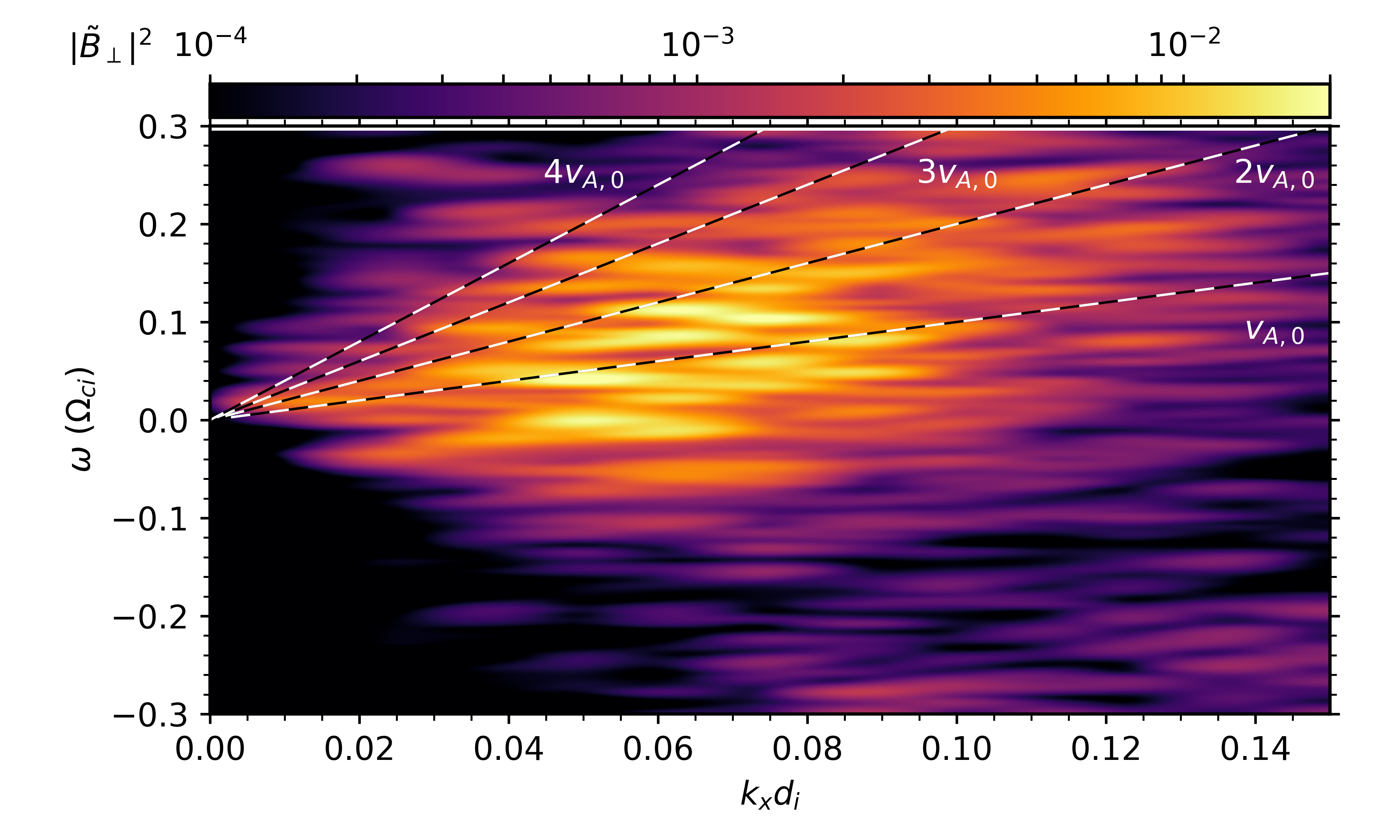}
\caption{Magnetic power spectrum (as in Figure \ref{fig:kw_up}) calculated $500 \di$ downstream of the shock, over the last $1200\ocii$ of the simulation. 
Most of the power is in modes traveling away from the shock (to the left, with $\omega>0$); the dashed lines indicate different negative integer values of $\w0$, between -1 and -4, consistent with Figure \ref{fig:alpha}.} \label{fig:kw_down}
\end{figure}

The Fourier analysis of the post-shock magnetic structures shown in Figure \ref{fig:kw_down} also supports this claim. 
The transform is calculated the same as in Figure \ref{fig:kw_up}, and over the last $1200\ocii$ of the simulation, in a window [$500\, \di,\ 200 \di$], positioned just downstream and moving with the shock;
the angular frequency versus wave number diagram is then boosted back into the downstream frame.
Most of the power is in modes with $\omega>0$, i.e., modes that move away from the shock with respect to the background plasma, at  speeds ranging between $-4\w0$ and $ -1\w0$, corresponding to the different dashed lines in Figure \ref{fig:kw_down} and consistent with the 2D time stack plots in Figure \ref{fig:Bxt}.

While the dispersion relationship is not as clear as Figure \ref{fig:kw_up}, due to the non-linear nature of the modes, we can still conclude that the postcursor magnetic structures preferentially move away from the shock, with phase speeds consistent with the Alfv\'enic and CR drift speeds in Figure \ref{fig:alpha}.
We point out that accounting for standard reflection/transmission of the  quasi-linear Alfv\'en waves observed upstream \citep{scholer+71} would return modes with $\omega< 0$ in the downstream, instead. 
Most likely, the failure of such a linear theory is due to the non-linear, collective, reaction of the shock transition itself to the hammering of the upstream fluctuations. 
The shock would therefore behave as an antenna that sends magnetic perturbations towards downstream, at a speed comparable to the ``natural'' speed for a magnetized plasma, namely the Alfv\'en speed\footnote{We thank S.~Schwartz for pointing out such an analogy.}.

From all of these diagnostics a picture arises where magnetic structures generated upstream, via CR-driven instabilities, are advected and compressed through the shock and retain their inertia over a sizable region downstream, forming the postcursor.
In the postcursor both the magnetic fluctuations and the CRs, which are isotropized in the wave frame, drift away from the shock faster than the thermal plasma, with a velocity on the order of the local Alfv\'en speed, $\w2$.
Since the post-shock magnetic pressure may become of order $\lesssim 10$ of the shock bulk ram pressure, such a drift is \emph{not negligible} with respect to the downstream fluid velocity;
for our benchmark run, $\w2\sim 0.5 u_2$.

This extra energy flux in both energetic particles and magnetic fields behind the shock affects the shock hydrodynamics at the zero-th order.
We present a discussion of the modifications to standard  Rankine-Hugoniot jump conditions necessary to account for this non-linear feedback in the next section.

\section{\label{sec:hydro}Modified Shock Hydrodynamics}
\subsection{Theory}
In the previous sections, we have illustrated the appearance of two non-linear features in the shock profile: the precursor and the postcursor. 
Here we address how they modify the shock compression and how energy is partitioned among different components;
for the effects that such modifications have on the spectrum of the accelerated particles, we refer to the companion paper, \cite{caprioli+20}.

Several works have developed a theory of CR-modified shocks including the dynamical role of energetic particles  \citep[e.g.,][]{drury83,jones+91,malkov+01,blasi02, amato+05,amato+06,malkov-volk96,caprioli+10c,caprioli12}, as well as of the self-generated magnetic field \citep[e.g.,][]{vladimirov+06,caprioli+08,caprioli+09a, slane+14}, but all of these studies only accounted for the CR precursor.
The main effect of the precursor is to produce a weaker subshock with compression ratio $\rs\equiv u_1/u_2<4$ and a total compression ratio $\rt\equiv u_0/u_2>4$.
The actual value of the total compression ratio depends on: 
1) the softer equation of state of CRs, which have an effective adiabatic index of $\gamma_c=4/3$; and 
2) the escape of CRs from far upstream, which makes the shock behave as partially radiative \citep[e.g.,][]{drury-volk81a,jones+91,caprioli+09b}.  
The former effect would saturate the total compression to 7, but with CR escape, $\rt$ may become very large.

To understand the role of the postcursor, we consider the conservation equations for mass, momentum, and energy in a 1D, non-relativistic, stationary, shock:
\begin{eqnarray}
    \left [ \rho u \right] = 0 \label{eq:jump0_n}\\
    \left [ \rho u^2 + P_{g} + P_{c} + P_{B} \right] = 0 \label{eq:jump0_p}\\
    \left [ \frac{1}{2}\rho u^3 + F_g + F_c +F_B \right] = 0\label{eq:jump0_E}
\end{eqnarray}
where $\gamma_i$, $P_i$, and $F_i$ are the adiabatic index, pressure, and energy flux of thermal gas, CRs, and magnetic fields ($i=g,c,B$, respectively).
We define the bulk flow velocity as ${\bf u}\equiv -u {\bf x}$, such that both $u > 0 $ and $F > 0$ even if ${\bf x}$ points from downstream to upstream, as in the simulations; 
the square brackets denote the difference between two arbitrary $x$ locations.
With appropriate prescriptions for $F_i$, this set of equations can be used to solve for the shock jump conditions.

The thermal gas energy flux has the canonical form:
\begin{equation}
    F_g (x)= \frac{\gamma_g}{\gamma_g -1}u(x)P_g(x)
\end{equation}
and the CR energy flux is linked to the CR pressure in a similar way, i.e.:
\begin{equation}
    F_c(x) = \frac{\gamma_c}{\gamma_c - 1} u_c(x) P_c(x),
\end{equation}
where $u_c$ is the CR bulk speed, including drifts.
Typically, most of the CR energy is in relativistic particles and $\gamma_c \approx 4/3$ but, for {\it ab-initio} simulations of non-relativistic shocks, non-relativistic CRs carry a sizable fraction of the energy \citep{haggerty+19a};
in our benchmark run, we measure $\gamma_c$ as the ratio of the enthalpy density to internal energy density and obtain $\gamma_c \approx 1.5$ (see Appendix \ref{ap:gamma} for more details).

The ``magnetic" energy flux, in general, has contributions from both the magnetic pressure and the kinetic energy associated with the plasma fluctuations, which for Alfv\'enic fluctuations reads \citep{scholer+71,vainio+99,caprioli+09a}:
\begin{equation}\label{eq:Bflux}
    F_B(x) = [2u_B(x) + u(x)]P_B(x),
\end{equation}
where $u_B= u \pm v_A$ is the local velocity of the magnetic fluctuations.
Equation \ref{eq:Bflux} encompasses the effective equation of state for the magnetic fluctuations, which, in general, cannot be expressed as polytropic; 
such an equation of state depends on the nature of the fluctuations, and Equation \ref{eq:Bflux} strictly holds only for Alfv\'enic perturbations.
In the absence of a general theory, we adopt the same prescription, even when non-resonant modes dominate, relying on the fact that in its nonlinear stage, Bell-driven turbulence looks quasi-Alfv\'enic \citep{riquelme+09,gargate+10,caprioli+14b}.

Using the results from Figure \ref{fig:kw_up} and \ref{fig:kw_down}, we have that in the precursor $u_{B}\simeq u-v_A$ and in the postcursor $u_{B}\simeq u+v_A$;
far upstream $u_{B,0} \simeq u_0$ because waves have no preferential direction. 

\begin{figure}[t]
    \includegraphics[width=.45\textwidth,clip=true,trim= 0 0 0 0]{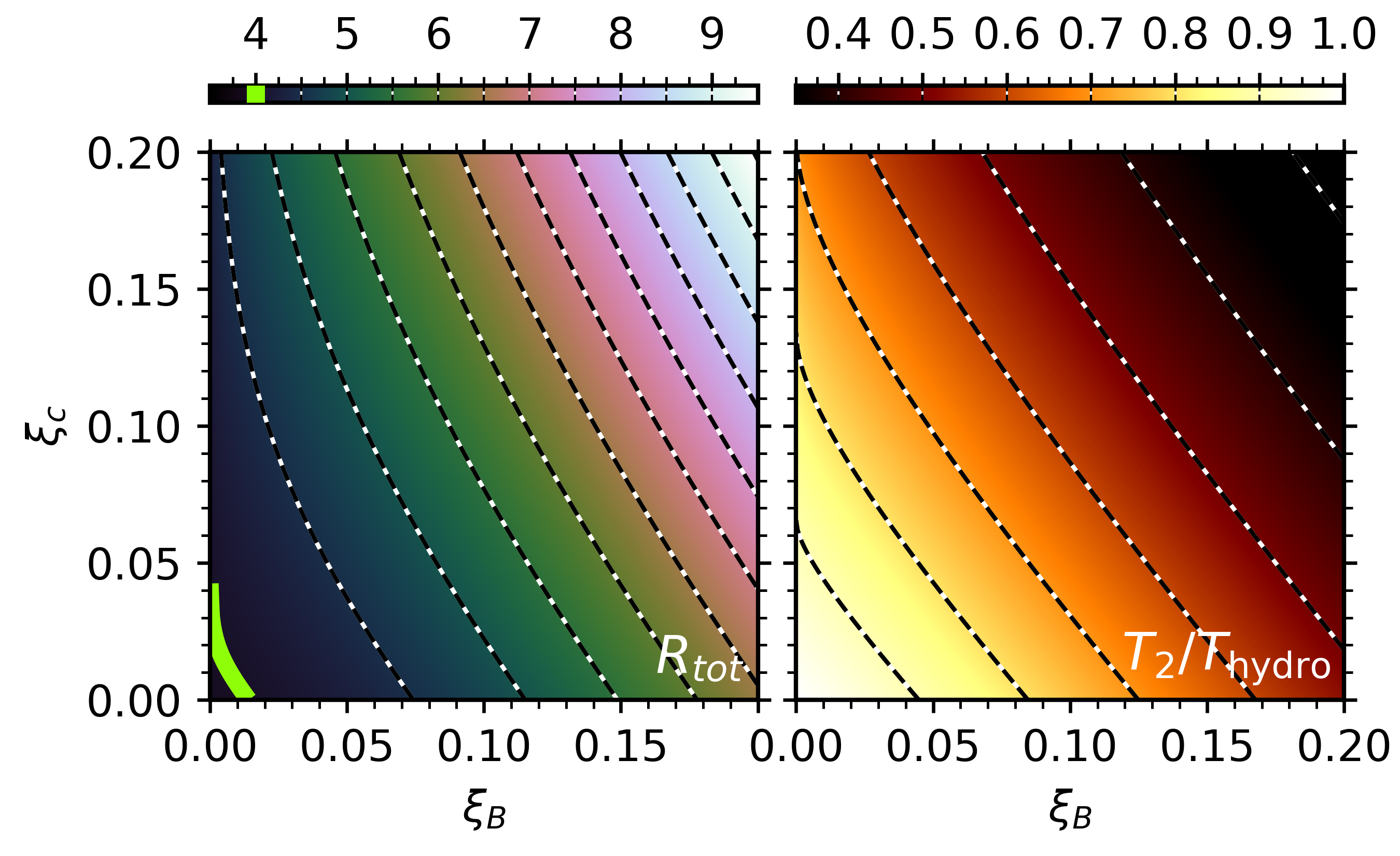}
\caption{Prediction from the CR-modified jump conditions (Equation \ref{eq:R}) for the total compression ratio ($\rt$, left panel) and the downstream temperature normalized to the unmodified prediction (right). 
Both are shown as a function of normalized downstream magnetic and CR  pressures, $\xi_B$ and $\xi_c$.
This solution is for $M_s \approx M_A \approx 20$, as in our benchmark simulation (\S\ref{sec:sims}), but generally holds for strong shocks.
The lime-green line marks the fiducial $\rt =4$ prediction for strong gaseous shocks.}
\label{fig:XiBr}
\end{figure}

Considering Equations \ref{eq:jump0_p} and \ref{eq:jump0_E} between 0 (far upstream) and 2 (postcursor), we normalize the momentum(energy) flux equation by dividing by the ram pressure (energy) $\rho_0u_0^2$,\ ($\rho_0u_0^3/2$), introduce the normalized pressure $\xi_i\equiv P_{i,2}/(\rho_0u_0^2)$  and  $\eta_i\equiv 2\gamma_i/(\gamma_i-1)$, obtaining:
\begin{equation}
    \xi_g \simeq  1 + \frac{1}{\gamma_g M_s^2} + \frac{1}{2 M_A^2}- \frac{1}{\rt} - \xi_{c} - \xi_B
    \label{eq:pressure}
\end{equation}
and
\begin{equation}
\begin{split}
    1 &+ \frac{\eta_g}{\gamma_g M_s^2} + \frac{3}{M_A^2} 
    \simeq \\
    & \frac{1}{\rt^2} + \frac{\eta_g\xi_g}{\rt} 
    + \frac{u_{c,2}}{u_2} \frac{\eta_c \xi_c}{\rt}
    + \left(\frac{2u_{B,2}}{u_2}+1\right) \frac{2\xi_B}{\rt},
    \label{eq:energy}
\end{split}
\end{equation}
where  $M_s \equiv \rho_0u_0^2/\gamma_g P_0$ and $M_A \equiv 4\pi \rho_0u_0^2/B_0^2$ are the far upstream sonic and Alfv\'enic Mach numbers, respectively.
Both the far upstream CR pressure and escape flux are negligible with respect to all the other terms for moderate $\xi_c$ \citep[see][for a detailed discussion]{caprioli+09b}, so we pose both $P_{c,0}\simeq 0$ and $F_{c,0}\simeq 0$.

As we saw in \S\ref{sec:post}, $u_{B,2}\simeq u_{c,2}\simeq u_2+\w2$ and hence:
\begin{equation}
    u_{c,2} \simeq u_{B,2} = u_2(1 + \sqrt{2\rt\xi_B}).
    \label{eq:alpha}
\end{equation}

Equations \ref{eq:pressure}, \ref{eq:energy}, and \ref{eq:alpha} can be rearranged into one equation for $\rt$ as a function of $M_s$, $M_A$ and the post-shock pressures $\xi_c$ and $\xi_B$. 
This equation ---Equation \ref{eq:R}, detailed in Appendix~\ref{ap:eq}--- is quartic with respect to $\sqrt{\rt}$ and its physical solution is shown in the left panel of Figure \ref{fig:XiBr}, where $\rt$ is plotted as a function of $\xi_B$  and $\xi_c$, for the values of $M_s$, $M_A$ and $\eta_c$ from our fiducial simulation. 
The striking conclusion from Figure \ref{fig:XiBr} is that --even if just a few percent of the pressure is channelled into CRs and magnetic fields-- $\rt$ becomes larger than the test-particle prediction of $4$ (light-green line). 
Note that, in contrast to the standard non-linear DSA theory without the postcursor, $\rt\gtrsim 4$ can be realized even if CRs are non-relativistic ($\gamma_c=5/3$) and do not escape upstream (see upper left subplot of Figure \ref{fig:ap_XiCB} in Appendix~\ref{ap:eq}).

In fact, on top of the usual compressibility enhancement due to the presence of both CRs and magnetic field, which have a softer equation of state than the thermal gas, a key role is played by the drift of the non-thermal components in the postcursor, which effectively acts as an energy sink.
Note that $\rt$ is more sensitive to $\xi_B$ than $\xi_c$, as the amplified magnetic field contributes to both the non-thermal energy downstream as well as the rate at which both CR and magnetic energy travel away from the shock.

These non-linear effects reduce the kinetic energy that is dissipated into heat at the shock. 
This can be seen in the right panel of Figure \ref{fig:XiBr}, which shows the fractional reduction in the downstream temperature, $T_2/T_2(\xi_B = \xi_c = 0)$, as a function of $\xi_B$ and $\xi_c$ for the values of $M_s$, $M_A$ and $\eta_c$ from our fiducial simulation. Again, even if relatively little energy is diverted to the CRs and magnetic fields, the predicted downstream temperature can decrease by as much as 50\%.

\subsection{Simulation Comparison}
Let us now compare the prediction for $\rt$ from the modified jump conditions with our benchmark simulation.
First of all, we need to find the (instantaneous) shock speed and boost all the velocities in that frame, which is the only stationary frame in which Equations \ref{eq:jump0_n} through \ref{eq:jump0_E} are valid;
this defines the actual value of $u_0 = \vup +|{\bf \vsh }|= \vup + u_2$ used for normalization purposes, i.e., 
\begin{equation}
    u_0= \frac{M\rt}{\rt -1} \w0.
\end{equation}
As in \S\ref{sec:post}, we calculate $\xi_c$ and $\xi_B$ by averaging over a region $500\di$ long behind the shock; 
the CR pressure is obtained by taking the appropriate moment of the ion distribution function above $\pinj$ in the downstream frame\footnote{This definition assumes that the CR distribution is isotropic in the downstream frame. This differs from the actual CR pressure on the order of $m_i\w2/p_c$ where $p_c$ is the momentum where most of the energy in the spectra resides, i.e., $p_c \sim \pinj$ for steep spectra and in our simulation $m_i\w2/p_c \sim 4\%$}: 
\begin{equation}
    \xi_c = \frac{4\pi}{3\rho_0u_0^2} \int_{\pinj}^\infty  v(p) f(p) p^3 dp.
\end{equation}

\begin{figure}
\includegraphics[width=.48\textwidth,clip=true,trim= 0 0 0 0]{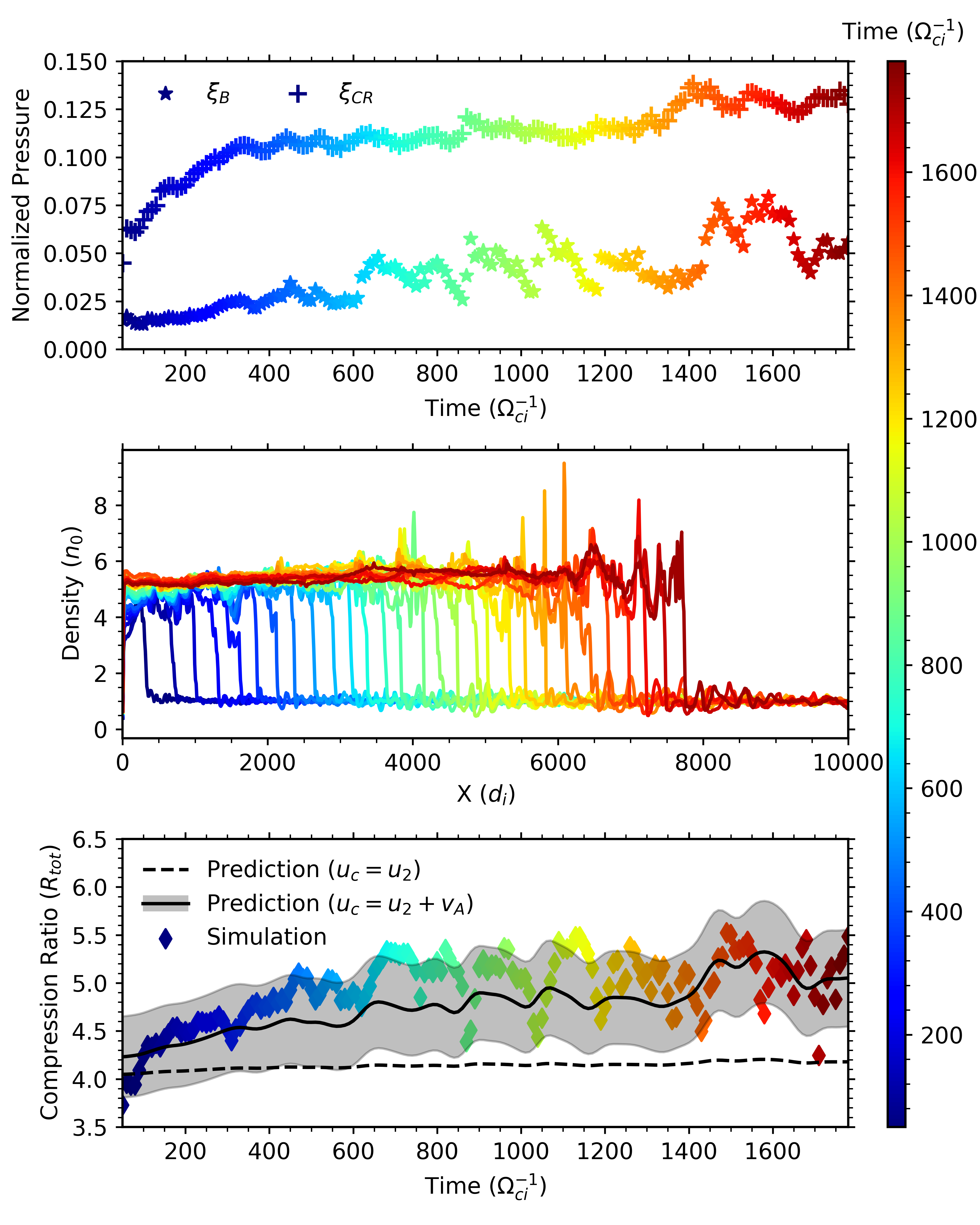}
\caption{
Time evolution (color coded) of physical quantities that show the CR-induced modification of our benchmark shock: 
postcursor normalized magnetic and CR pressures, $\xi_B$ and $\xi_c$  (top panel), density profile (middle panel), and total compression ratio, $\rt$ (bottom panel).
The CR pressure quickly converges to $\xi_c\approx 10\%$, while $\xi_B$ saturates around 6\%;
at the same time, the compression ratio departs from the test-particle value of $\sim 4$ and becomes as large as $\rt\gtrsim 5.5$.
The prediction including the postcursor drift (Equation \ref{eq:R}) is shown as a solid line with an error of 10\% (grey band) and fits the simulation much better than the standard CR-modified shock prediction (dashed line).}\label{fig:R}
\end{figure}

$\xi_c$ and $\xi_B$ are plotted as a function of time in the first panel of Figure \ref{fig:R} (crosses and stars, respectively);
the color code corresponds to the time in the simulation.
Together, the normalized CR and magnetic pressure make up about 15-20\% of the pressure budget in the postcursor region.
$\xi_c$ increases quickly to a value just above $0.1$ and remains nearly constant throughout the simulation, whereas the magnetic pressure rises more slowly up to a value of $0.05-0.075$ towards the end of the simulation.
The order of magnitude of the normalized pressures in the simulation makes it clear that, based on the predictions from the modified jump condition in Equation \ref{eq:R}, the total compression ratio should be larger than 4.
The second panel of Figure \ref{fig:R} shows the $y$-averaged density profile as a function of time:  
at early times (blue lines) the compression ratio $\rt\approx 4$, but it increases with time up to $\rt\gtrsim 5 - 6$ toward the end of the simulation (red lines).

The agreement between the prediction based on the modified jump conditions and the simulation can be quantified by determining a time-dependent $\rt(t)$, averaged over the postcursor.
The solution of Equation \ref{eq:R} for the actual values of $\xi_c(t)$ and $\xi_B(t)$ (top panel) is compared with the measured value of $\rt$ (diamonds) in the bottom panel of Figure \ref{fig:R}; the grey-shaded area corresponds to a fiducial 10\% error on the prediction that encompasses the uncertainty in measuring pressures and velocities in a profile with small-scale spatial variation, as well as the assumption of Alfv\'enic-like magnetic turbulence (Equation \ref{eq:Bflux}). 
We find a general agreement between theory and simulations in the value, trend, and periodic variation of $\rt$, with minor deviations that can potentially be attributed to time evolution (not captured by Equations \ref{eq:jump0_n}-\ref{eq:jump0_E}) and transient shock features.
To stress the importance of the postcursor in the shock dynamics, the bottom panel of Figure \ref{fig:R} includes as a dashed line the prediction with no CR/magnetic drift ($u_c = u_b = u_2$).
As mentioned above, CR and magnetic pressure terms alone are not sufficient to account for the strong shock modification that we observe. 

The predicted values of $\rt$ can also account for the quasi-periodic behaviour of the shock driven by the variations in $\xi_b$ (top panel of Figure \ref{fig:R}).
Such oscillations hinge on the same precursor physics discussed in Section \ref{sec:pre} and will be explored in greater detail in future works.

\subsection{A Critical Review of Previous Results}
Such modifications on the shock hydrodynamics  were not seen in previous hybrid campaigns that covered similar parameter space \citep{caprioli+14a,caprioli+14b} because of the choice of the effective electron polytropic index, $\gamma_e$.
When an effective $\gamma_e$ is chosen in order to mimic temperature equilibration and pressure balance between post-shock electrons and ions, a large value $\gamma_e\approx 3-4$ must be chosen for strong shocks \citep[see][for details]{gargate+07,caprioli+18}. 
However, enforcing electron-ion equilibration via an effective $\gamma_e$ requires assuming \emph{a priori} the value of the realized compression ratio; 
therefore, if one assumes that $\rt\approx 4$, the artificially-stiff electron equation of state ends up limiting the total compression that can be achieved. 
Using such a prescription, \cite{caprioli+14a} found $\rt\sim 4.4$, underestimating the shock modification that we report here with the adiabatic equation of state.
Note that as long as the post-shock pressure does not become \emph{dominated} by electrons (which is unphysical but may happen for $\gamma_e\gg5/3$), the shock modification does not depend on whether electrons are adiabatic or in equipartition with the ions.

Very recently \cite{bret20}  pointed out how standard MHD jump conditions are not commonly realized in many papers based on PIC simulations of shocks; 
our formalism, in which the kinetic backreaction of accelerated particles and self-generated magnetic fields is accounted for, naturally presents itself with a physically-motivated framework in which to study the agreement between simulations and theory.

\subsection{Mach Number Dependence}\label{sec:Mach}
\begin{figure}[t]
\includegraphics[width=.48\textwidth,clip=true,trim= 0 0 0 0]{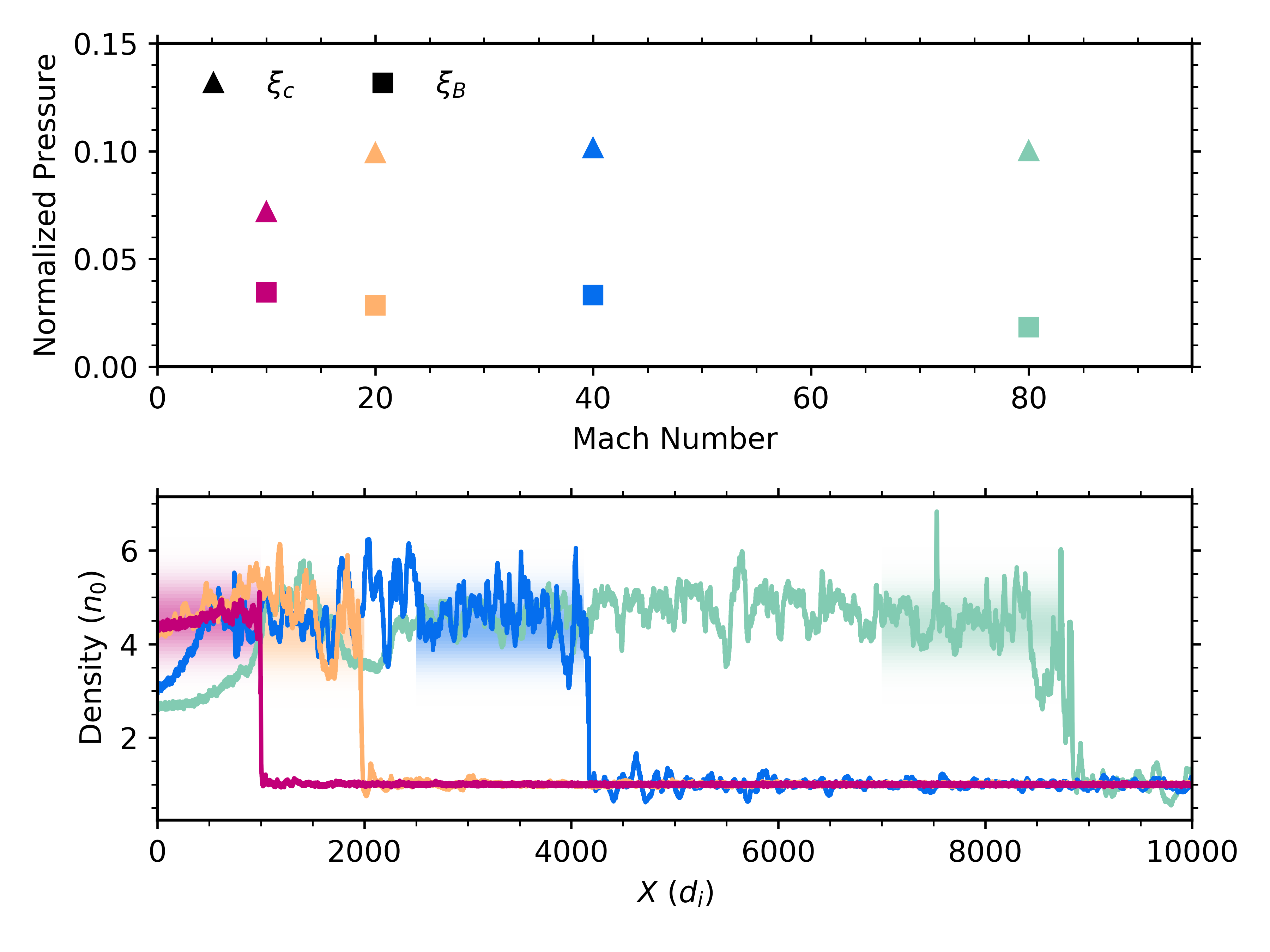}
\caption{Top panel: normalized CR ($\xi_c$, triangles) and magnetic pressure ($\xi_B$, squares) for 4 different shock Mach numbers (color-coded). 
These values are used in Equation \ref{eq:R} to predict the modified $\rt$ (shaded colored regions), which are compared with 
the actual density profiles for different $M$ taken at $t=375 \ocii$ for each simulation (bottom panel). 
For all these strong shocks $\rt$ increases beyond 4, in agreement with the theory outlined here.}\label{fig:M}
\end{figure}
Beyond our benchmark $M = 20$ simulation, we have run parallel shocks with $M=10,\, 40$ and $80$, and the results are shown in Figure \ref{fig:M}.
As already reported by \cite{caprioli+14a}, the normalized CR pressure ---a proxy for acceleration efficiency--- is generally about 10\%; 
the normalized magnetic pressure, instead, is typically 2-5\%, values commonly inferred in multi-wavelength analysis of SNRs \citep[e.g.,][]{volk+05,parizot+06,caprioli+08}.

Consistent with our predictions, the shocks in each of these simulations have compression ratios that exceed the gaseous value. 
This enhancement is shown in the bottom panel of Figure \ref{fig:M}, in which the $y$-averaged density profiles are shown. 
The color-coded shaded regions correspond to the prediction of $\rt$ from Equation \ref{eq:R} and are in good agreement with the simulated values for each Mach number, which strengthens the applicability of these results to many heliospheric/astrophysical systems.

At larger Mach numbers $M\gtrsim 30$ the magnetic field amplification in the precursor is controlled by the growth of the Bell instability  \citep[e.g.,][]{amato+09}, whose fastest growing modes are purely growing (i.e., with almost zero phase speed) and with wavelengths much shorter than the CR gyroradii.
Nevertheless, global hybrid simulations of shocks with $M=60,80$ and 100 have shown that small-wavelength modes saturate rather quickly far upstream, and that in the precursor most of the power in self-generated fields is still in modes quasi-resonant with the accelerated particles \citep{caprioli+14b,caprioli+14c}.
In the non-linear stage of the Bell instability waves start propagating with a phase speed close to the Alfv\'en speed in the amplified field \citep{riquelme+09,gargate+10}, so we expect that the general phenomenology outlined in this work should apply also at very strong shocks, such as those in SNRs.

\section{\label{sec:disc}Conclusions}
In this work we use self-consistent hybrid simulations to study the modifications that self-generated CRs and associated magnetic turbulence induce on the dynamics of a collisionless plasma shock.
The efficient acceleration of CRs leads to a pre-compression and deceleration of the plasma upstream of the shock: 
in such a precursor region, there is non-adiabatic heating of the inflowing plasma, likely a byproduct of the CR-driven magnetic fluctuations \citep{caprioli+14a}. 
Additionally, we identify quasi-periodic fluctuations in the magnetic field strength and CR pressure in the precursor, which are attributed to the shock geometry transitioning back and forth between quasi-parallel and oblique/quasi-perpendicular configurations.

Here, we report for the first time the formation of a characteristic region downstream of the shock, which we call the \emph{postcursor}, where the enhanced magnetic fluctuations generated upstream and then compressed by the shock play a crucial dynamical role.
Such magnetic fluctuations are found to propagate away from the shock in the downstream rest frame, with a velocity comparable to the local Alfv\'en speed in the amplified magnetic field (Figures \ref{fig:Bxt} and \ref{fig:kw_down}).
In the postcursor, CRs become isotropic in a frame moving with the magnetic fluctuations, rather than in the downstream fluid frame, resulting in a peculiar drift with respect to the background plasma (Figure \ref{fig:alpha}).

Such non-linear features, ultimately driven by CR physics, lead to an enhanced shock compression ratio.
More precisely, the total compression ratio becomes larger than the standard value of $\rt=4$, due, not only to the compressibility of relativistic CRs and magnetic fields, but mainly because of the larger rate at which the non-thermal populations are advected away from the shock.
The solution of the modified jump conditions (Equation \ref{eq:R}) is presented and compared with simulations proving general agreement between the predicted and measured compression ratio as a function of time (Figure \ref{fig:R}).
Even a moderate CR acceleration efficiency, $\xi_c\sim 10\%$, is able to increase the shock compression ratio for a large Mach number shock by nearly 50\% from the standard fluid prediction to $\rt \sim 5.5$.

We have tested that this behaviour is not limited to our benchmark case by running simulations with $M=10,\, 40$, and $80$, which present a very similar phenomenology. 
While it is computationally very challenging to run kinetic simulations of strong shocks for much longer than we did, we do not see strong evolution of the shock modification in the last few hundreds $\ocii$ and we achieve quite large values for $\xi_c$ and $\xi_B$, which may suggest that in realistic shocks the compression ratio should not be much greater than $\rt\gtrsim 6$. 
Provided that a reliable prescription for CR injection were implemented, the long-term shock evolution could be followed with hybrid+MHD codes, in which thermal particles are described as a fluid \citep{zachary+86,lb00,reville+12,bai+15,vanmarle+18}.

The enhanced compression ratios found in these hybrid simulations can be regarded as the first \emph{ab-initio} evidence of the existence of CR-modified shocks, which had been suggested almost 40 years ago \citep[][]{drury-volk81a,drury-volk81b} but never verified in kinetic simulations.
Observational hints of shock compression ratios larger than 4 have been reported for young supernova remnants, such as Tycho  \citep{warren+05} and SN1006 \citep{gamil+08};
in particular, in SN1006 the distance between the forward shock and the contact discontinuity is inferred to be modulated with the azimuth, being smaller (corresponding a larger compression ratio) where the shock is quasi-parallel \citep[see][]{reynoso+13}, i.e., the region where CR acceleration is expected to be more prominent \citep{caprioli+14a,caprioli15p}.

The values of $\xi_c\sim 5-15\%$ and $\xi_b\sim 2-10\%$ required to produce $\rt\gtrsim 4$ are  consistent with the values inferred from multi-wavelength observations of young SNRs \citep[e.g.,][]{volk+05,parizot+06,caprioli+08,morlino+12,slane+14}, so we expect CR-induced shock modification to be a ubiquitous phenomenon. 

An important result that follows from our simulations is that the main driver of the shock modification, i.e., the magnetic drift in the postcursor becoming comparable to the local Alfv\'en speed, can be inferred observationally.
In fact, the postshock magnetic field can be constrained with non-thermal X-rays \citep[e.g.,][]{bamba+05,ballet06,uchiyama+07,morlino+10,ressler+14,tran+15} and the gas density and temperature by X-ray \citep[e.g.,][]{warren+05,miceli+12, slane+14} and Balmer line emission \citep[e.g.,][]{chevalier+78,ghavamian+07,blasi+12,morlino+13,knezevic+17}, while the shock speed can be estimated from proper motion of X-ray and/or Balmer features.
In principle, high-resolution X-ray observations can even test the presented theory as a function of the local shock inclination, e.g., in bilateral SNRs such as SN1006, probing whether CR-modified shocks manifest themselves in quasi-parallel regions.
A corollary of our results is that quasi-perpendicular shocks, which are generally poor ion injectors \citep{caprioli+15,caprioli+18}, should not exhibit deviations from standard Rankine--Hugoniot conditions. 

A natural question that arises is {\it what is the spectrum of the particles accelerated in a CR-modified shock?} 
For a dedicated discussion of such a critical question we refer to a companion paper, \cite{caprioli+20}.

\software{\dHybridR~\citep{haggerty+19a}, 
}

\acknowledgments
We would like to thank  P.~Blasi, E.~Amato, A.~Spitkovsky, D.~Eichler, L.~O'C. Drury, S.~Schwartz, and L.~Wilson III for stimulating and constructive discussions. 
This research was partially supported by NASA (grant NNX17AG30G, 80NSSC18K1218, and 80NSSC18K1726), NSF (grants AST-1714658, AST-1909778, PHY-1748958, PHY-2010240), and by the International Space Science Institute’s (ISSI) International Teams program.
Simulations were performed on computational resources provided by the University of Chicago Research Computing Center, the NASA High-End Computing Program through the NASA Advanced Supercomputing Division at Ames Research Center, and XSEDE TACC (TG-AST180008).

\appendix
\section{Simulation Details}\label{ap:sims}
In addition to the the benchmark $M = 20$ simulation discussed in Section~\ref{sec:sims}, five simulations were performed to control for Mach number and box width. The parameters of each simulation are given in Table~1 and from left to right are: Simulation ID, Mach number ($M$), upstream ion plasma beta ($\beta_i$), simulation speed of light ($c$), grid size ($\Delta x$), time step ($\Delta t$), simulation length ($L_x$) and simulation width ($L_y$). Each simulation uses 16 macro-particles per $\di^2$. 
The reduction in the box length for  M80 was to compensate for the added computational cost of the reduced time step.
\begin{center}
    
\noindent\begin{minipage}{.7\textwidth}
    \centering 
 \begin{tabular}{|c | c | c | c | c | c | c | c|} 
 \hline
 Sim & $M$ & $\beta_i$ & $c\ (\w0)$ & $\Delta x\ (\di)$ & $\Delta t\ (\ocii)$ 
     & $L_x\, (d_i)$ & $L_y\, (d_i)$ \\
 \hline
 M10 & 10 & 2 & 50 & 0.5 & 0.005 & $10^5$ & 200\\
 M20 & 20 & 2 & 100 & 0.5 & 0.0025 & $10^5$ & 200\\
 M20w & 20 & 2 & 100 & 0.5 & 0.0025 & $10^5$ & 1000\\
 M40 & 40 & 2 & 200 & 0.5 & 0.00125 & $10^5$ & 200\\
 M80 & 80 & 2 & 400 & 0.5 & 0.000625 & $5\times 10^4$ & 200\\
 \hline
\end{tabular} 
\end{minipage}%
\end{center}

\section{The Solution to the CR-Modified Jump Conditions}\label{ap:eq}
The total compression ratio can be found by combining equations \ref{eq:pressure}, \ref{eq:energy}, and \ref{eq:alpha}. The equations can be rewritten into a single quartic equation in terms of $X=\sqrt{\rt}$ where the coefficients depend on $M_s$, $M_A$ and the post-shock pressures $\xi_c$ and $\xi_B$, namely:
\begin{align}\label{eq:R}
    c_1 &X^{4} + c_2 X^{3} + c_3 X^2 + c_4 = 0;
\end{align}
with the coefficients:
\begin{align*}
    \begin{split}
    c_1 &= 1 + \frac{\eta_g}{\gamma_g M_s^2} + \frac{3}{M_A^2}\\
    c_2 &= -\sqrt{2\xi_B}\left(\eta_c\xi_c + 4\xi_B\right)\\
    c_3 &= \eta_g(\xi_c + \xi_B - 1 - \frac{1}{\gamma_g M_s^2} - \frac{1}{2M_A^2}) - \eta_c\xi_c - 6\xi_B\\
    c_4 &= \eta_g - 1.
    \end{split}
\end{align*}
Since $\gamma_g=5/3$, $\eta_g =5$, while for $\gamma_c$ between 5/3 and 4/3, $\eta_c$ varies between 5 and 8;
in our benchmark simulation, $\gamma_c\simeq 1.5$ and $\eta_c\simeq 6$ (see Appendix~\ref{ap:gamma}), which is also used for the predictions for different Mach number simulations in Figure \ref{fig:M}.
Note that it is not possible to introduce an effective adiabatic index for the wave/magnetic fields because the relationship between pressure and energy density is non trivial even when assuming Alfv\'en waves (Equation \ref{eq:Bflux}).

Equation \ref{eq:R} has 4 roots, but only 1 of which is physically relevant. 
Since $\xi_c,\xi_B\ll 1$, two of the roots are negative and can be neglected. 
Of the two positive remaining roots, only one corresponds to an increase in density, temperature and entropy at the shock and thus is the physical solution. 
Note that for non-null $\xi_c$ and $\xi_B$, the trivial solution $X=1$ disappears.

\section{The CR Adiabatic Index}\label{ap:gamma}
\begin{figure} 
\centering 
\includegraphics[width=.48\textwidth,clip=true,trim= 0 0 0 0]{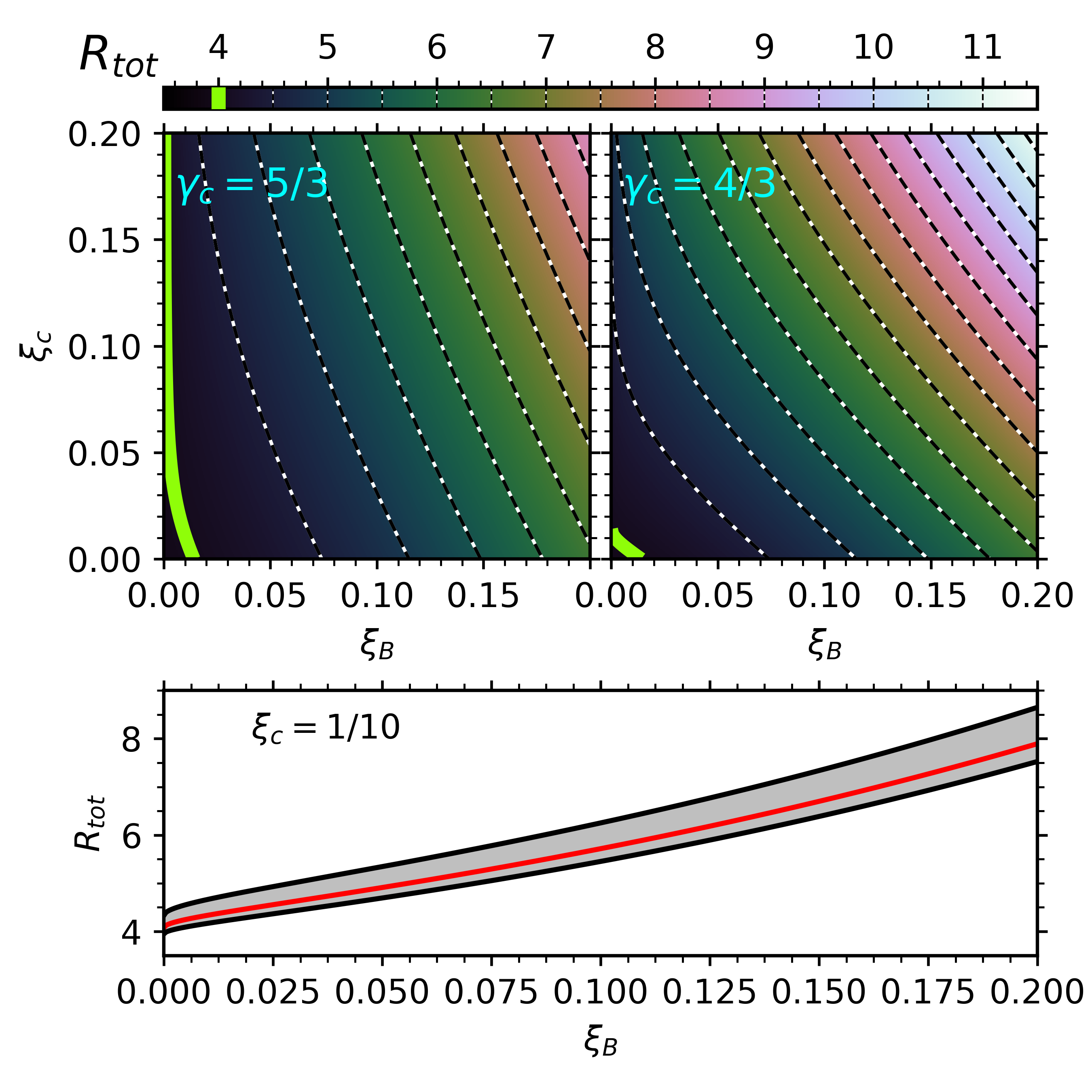}
\caption{Top panels: total compression ratio, $\rt$, calculated from the modified jump conditions (Equation \ref{eq:R}) for non-relativistic ($\gamma_c = 5/3$, left) and ultra-relativistic CRs ($\gamma_c = 4/3$, right). 
Both are calculated with $M_s \approx M_A \approx 20$ as in our benchmark  simulation and shown as a function of normalized downstream magnetic and CR pressures, $\xi_B$ and $\xi_c$.
The lime-green line marks the fiducial $\rt =4$ prediction for strong gaseous shocks.
A 1D cut of these solutions is shown in the bottom panel, varying $\xi_B$ for constant $\xi_c = 0.1$. The shaded region is bound by the non-relativistic and ultra-relativistic cases, and the solution using the measured $\gamma_c = 1.5$ is shown by the red line.}\label{fig:ap_XiCB}
\end{figure}

To correctly predict the hydrodynamic modifications, an effective adiabatic index, $\gamma_c$, must be determined for the transrelativistic CR distribution.
$\gamma_c$ is measured as the ratio of the enthalpy density to the internal energy density:
\begin{equation}
    \gamma_c = 1 + \frac{\int_{p_{inj}}^\infty vp^3/3f dp}{\int_{p_{inj}}^\infty m_ic^2(\Gamma - 1)p^2f dp}
\end{equation}
where $\Gamma$ is the Lorentz factor.
For non-relativistic particles, $mc^2(\Gamma - 1) \approx pv/2$ and  $\gamma_c = 5/3$, while for relativistic distributions, $mc^2(\Gamma - 1) \approx pc$ yielding $\gamma_c = 4/3$. 
In general, the exact value for the $\gamma_c$ will depend on the CR distribution function, but will be bounded by these two values.
The top panels of Figure \ref{fig:ap_XiCB} show the upper and lower bounds for the solution for $\rt$ based on Equation \ref{eq:R}. 
In both the non-relativistic (left) and relativistic (right) cases, we obtain $\rt\gtrsim 4$ even for modest values of $\xi_B$ and $\xi_c$; in general, $\rt$ is larger in the relativistic case.
To further illustrate the effect of the CR equation of state, the predicted compression ratio for a fixed value of $\xi_c=0.1$ and varying $\xi_B$ are shown in the bottom panel of Figure \ref{fig:ap_XiCB};
varying the CR adiabatic index modifies $\rt$ by about 10\% at most, an effect less important than the one induced by the CR drift in the precursor, which is controlled by $\xi_B$.

\bibliography{Total}

\end{document}